\documentclass[10pt,journal]{IEEEtran}
\usepackage{lipsum}
\usepackage{amssymb}
\usepackage{pifont}
\usepackage{color}

\usepackage{float}
\usepackage[caption = false]{subfig}
\usepackage[final]{graphicx}
\usepackage{hyperref}

\ifCLASSINFOpdf
\else
\fi
\begin{document}
%
% paper title
% Titles are generally capitalized except for words such as a, an, and, as,
% at, but, by, for, in, nor, of, on, or, the, to and up, which are usually
% not capitalized unless they are the first or last word of the title.
% Linebreaks \\ can be used within to get better formatting as desired.
% Do not put math or special symbols in the title.
\title{Deep Learning for Encrypted Traffic Classification: An Overview}
%
%
% author names and IEEE memberships
% note positions of commas and nonbreaking spaces ( ~ ) LaTeX will not break
% a structure at a ~ so this keeps an author's name from being broken across
% two lines.
% use \thanks{} to gain access to the first footnote area
% a separate \thanks must be used for each paragraph as LaTeX2e's \thanks
% was not built to handle multiple paragraphs
%

\author{Shahbaz~Rezaei,~\IEEEmembership{Member,~IEEE,}
        and~Xin~Liu,~\IEEEmembership{Senior,~Member,}% <-this % stops a space
\thanks{S. Rezaei and X. Liu are with Computer Science Department,
University of California, Davis, USA (e-mails: srezaei@ucdavis.edu
and liu@cs.ucdavis.edu).}% <-this % stops a space
\thanks{
© 2019 IEEE. Personal use of this material is permitted. Permission
from IEEE must be obtained for all other uses, in any current or future
media, including reprinting/republishing this material for advertising or
promotional purposes, creating new collective works, for resale or
redistribution to servers or lists, or reuse of any copyrighted
component of this work in other works.”
}}
%\thanks{Manuscript received April 19, 2005; revised August 26, 2015.}}

% note the % following the last \IEEEmembership and also \thanks - 
% these prevent an unwanted space from occurring between the last author name
% and the end of the author line. i.e., if you had this:
% 
% \author{....lastname \thanks{...} \thanks{...} }
%                     ^------------^------------^----Do not want these spaces!
%
% a space would be appended to the last name and could cause every name on that
% line to be shifted left slightly. This is one of those "LaTeX things". For
% instance, "\textbf{A} \textbf{B}" will typeset as "A B" not "AB". To get
% "AB" then you have to do: "\textbf{A}\textbf{B}"
% \thanks is no different in this regard, so shield the last } of each \thanks
% that ends a line with a % and do not let a space in before the next \thanks.
% Spaces after \IEEEmembership other than the last one are OK (and needed) as
% you are supposed to have spaces between the names. For what it is worth,
% this is a minor point as most people would not even notice if the said evil
% space somehow managed to creep in.

% The paper headers
\markboth{Submitted to IEEE Communications Magazine}%
{Shell \MakeLowercase{\textit{et al.}}:  Deep Learning for Encrypted Network Traffic Classification: An Overview}
% The only time the second header will appear is for the odd numbered pages
% after the title page when using the twoside option.
% 
% *** Note that you probably will NOT want to include the author's ***
% *** name in the headers of peer review papers.                   ***
% You can use \ifCLASSOPTIONpeerreview for conditional compilation here if
% you desire.

% If you want to put a publisher's ID mark on the page you can do it like
% this:
%\IEEEpubid{0000--0000/00\$00.00~\copyright~2015 IEEE}
% Remember, if you use this you must call \IEEEpubidadjcol in the second
% column for its text to clear the IEEEpubid mark.

% use for special paper notices
%\IEEEspecialpapernotice{(Invited Paper)}

% make the title area
\maketitle

% As a general rule, do not put math, special symbols or citations
% in the abstract or keywords.
\begin{abstract}
Traffic classification has been studied for two decades and applied to a wide range of applications from QoS provisioning and billing in ISPs to security-related applications in firewalls and intrusion detection systems. Port-based, data packet inspection, and classical machine learning methods have been used extensively in the past, but their accuracy have been declined due to the dramatic changes in the Internet traffic, particularly the increase in encrypted traffic. With the proliferation of deep learning methods, researchers have recently investigated these methods for traffic classification task and reported high accuracy. In this article, we introduce a general framework for deep-learning-based traffic classification. We present commonly used deep learning methods and their application in traffic classification tasks. Then, we discuss open problems and their challenges, as well as opportunities for traffic classification.
\end{abstract}
% Note that keywords are not normally used for peerreview papers.

% For peer review papers, you can put extra information on the cover
% page as needed:
% \ifCLASSOPTIONpeerreview
% \begin{center} \bfseries EDICS Category: 3-BBND \end{center}
% \fi
%
% For peerreview papers, this IEEEtran command inserts a page break and
% creates the second title. It will be ignored for other modes.
\IEEEpeerreviewmaketitle

\section{Introduction}
% The very first letter is a 2 line initial drop letter followed
% by the rest of the first word in caps.
% 
% form to use if the first word consists of a single letter:
% \IEEEPARstart{A}{demo} file is ....
% 
% form to use if you need the single drop letter followed by
% normal text (unknown if ever used by the IEEE):
% \IEEEPARstart{A}{}demo file is ....
% 
% Some journals put the first two words in caps:
% \IEEEPARstart{T}{his demo} file is ....
% 
% Here we have the typical use of a "T" for an initial drop letter
% and "HIS" in caps to complete the first word.
\IEEEPARstart{T}{raffic} classification, the categorization of network traffic into appropriate classes, is important to many applications, such as quality of service (QoS) control, pricing, resource usage planning, malware detection,  and intrusion detection. Because of its importance, many different approaches have been developed over years to accommodate the diverse and changing needs of different application scenarios. In particular, new advances in communications, including encryption and port obfuscation, raise additional challenges to  network classification.

Traffic classification techniques have evolved significantly over time. The first and easiest approach is to use port numbers. However, its accuracy has been decreasing 
%port-number-based approach is not as accurate as it used to be 
because newer  applications either use well-known port numbers to disguise their traffic or do not use standard registered port numbers. Despite its inaccuracy, the port number is still widely used either alone or in tandem with other features in practice. 
%whenever the accuracy is not critical.
%The next generation of traffic classifiers, called payload-based or data packet inspection (DPI) methods, focused on finding patterns or keywords in packet content. Payload-based methods were only useful at the nascent stage of their introduction when most Internet traffic was not encrypted. Moreover, inspecting the entire payload is a slow and computationally costly process to be used in an operational network with high bandwidth links. As a result, a new generation of methods, called flow statistics-based, emerged. These methods rely on statistical or time series features for classification enabling them to deal with encrypted traffic. These methods usually employed classical machine learning algorithms, such as support vector machine (SVM), random forest (RF) and k-nearest neighbor (KNN), and achieved better performance in specific task they are designed for. However, the performance of these methods heavily depend on the features selected carefully by a domain expert which limits their generalizability.
The next generation of traffic classifiers, relying on payload  or data packet inspection (DPI), focuses on finding patterns or keywords in data packets. These methods are only applicable to unencrypted traffic and has high computational overhead. As a result, a new generation of methods, based on flow-statistics, emerged. These methods rely on statistical or time series features, which enable them to handle both encrypted and unencrypted traffic. These methods usually employ classical machine learning (ML) algorithms, such as random forest (RF) and k-nearest neighbor (KNN). However, their performance heavily depends on the human-engineered features, which limit their generalizability.

Deep learning obviates the need to select features by a domain expert because it automatically selects features through training. This characteristic makes deep learning  a highly desirable approach for traffic classification, especially when new classes constantly emerge and patterns of old classes evolve. 
% because .... Furthermore,   
%employs hierarchical feature selection. 
%A deep learning model can be periodically retrained with new data to adopt to new changes, and thus is inherently suitable for traffic classification tasks where
Another important characteristic of deep learning is that it has a considerably higher capacity of learning in comparison to traditional ML methods, and thus can learn highly complicated patterns. Combining these two characteristics,  as an end-to-end approach, deep learning is capable of learning the non-linear relationship between the raw input and corresponding output without the need to break the problem into the small sub-problems of feature selection and classification.

%Recently, a few studies have shown the efficacy of simple deep learning methods on traffic classification task. However, deep learning have advanced significantly in the past few years and many useful methods and techniques have been introduced to tackle classical machine learning problems, such as natural language processing and computer vision, which have not been employed for traffic classification yet.

Recent work has demonstrated the efficacy of deep learning methods in traffic classification, in particular, in encrypted traffic. To achieve this goal, DL requires sufficient labeled data and adequate computation power. In this article, we will overview the general framework for (encrypted) traffic classification task. We provide general guidelines for classification tasks, including data collection and cleaning, feature selection, and model selection. Moreover, we discuss deep learning techniques and how they have been applied for traffic classification task. Finally, open problems and future directions are discussed.

\section{Overview of Classification Problems on Computer Network}

\begin{figure*}
\centering
\includegraphics[width = .8\linewidth]{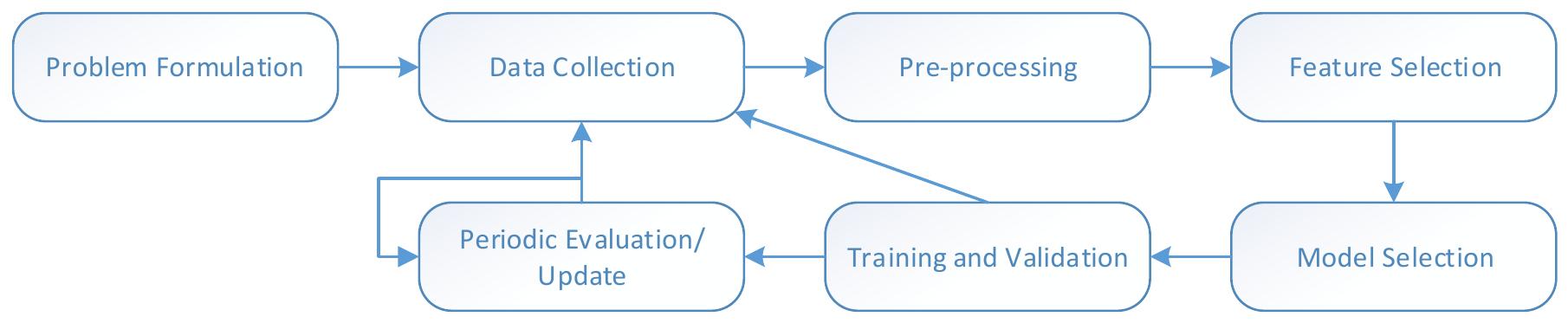}
\caption{General Framework to build a network classifier.}
\label{fig-framework}
\end{figure*}

%Several papers have already studied network-related classification problems using classical machine learning algorithms or deep learning methods. 
Fig. \ref{fig-framework} illustrates a general framework for traffic classification, comprising seven steps. Most existing work adopts all or part of the framework. We discuss the first four steps in this section, and the last three in the next section, with a focus on deep-learning-based approaches.
%For cases where capturing and labeling enough data is possible, and also memory and computational budget is not extremely limiting, a general framework consisting of seven steps, as shown in Fig. \ref{fig-framework}, has shown to lead to an acceptable classifier. In this section, we discuss the first four steps in our general framework in details. The last thee steps will be discussed in next sections.

\subsection{Problem Definition}

The first step to build a network traffic classifier is to clearly define the goal of classification. Typical goals include QoS provisioning, resource usage planning, billing system customization, intrusion detection, and malware detection.
To serve its goal, one can categorize traffic classes based on 1) protocols (e.g. UDP, TCP, FTP or HTTP), 2) applications (e.g. Skype, WeChat or Torrent), 3) traffic-types (e.g. browsing, downloading or video chat), 4) websites, 5) user actions (e.g. posting a comment or sending voice message), 6) operating systems, 7) browsers, and so on.
Hence, the goal is to label each flow with corresponding traffic classes. A flow is usually determined by a 5-tuple: Source IP, destination IP, source port, destination port and protocol.

Furthermore, traffic classification can also be categorized into  two sub-classes: online and offline. Online classification usually refers to the cases where flows need to be classified as fast as possible, usually within the first few to tens of packets. For instance, for QoS provisioning and routing,  classification needs to be online because the output of the classification is directly used for decisions on the current flow. For other applications, such as billing systems,  classification can be offline. 

While traffic classification applies to vastly different scenarios, most studies share two ubiquitous aspects: (a) the input data for classification is raw packet data, part of it, or information directly derived from it, (b) Similar ML algorithms are used. The focus of the article is on encrypted application/traffic-type classification. Yet, the same methodology may be used for other classification problems with minor modifications. 

\subsection{Data Collection}
One of the most important requirements for training a deep learning model is a large and representative dataset. Although there are a few public and recent datasets available for research purpose\footnote{
https://wand.net.nz/wits/ [Accessed on 1/15/2019]

https://projects.cs.dal.ca/projectx/Download.html [Accessed on 1/15/2019]

 http://www.unb.ca/cic/datasets/vpn.html [Accessed on 1/15/2019]
},  
there is no commonly agreed-upon dataset for most traffic-related classification problems. Possible reasons include 1) the number of possible traffic classes is enormous and it is practically impossible for one dataset to contain all traffic types, 2) there is no commonly accepted data collection and labeling methods, 3) different collection methods and scenarios result in different feature availability and distributions.

In practice, researchers often collect a dataset specific to their classification goal. To do so, the first step is to determine a data collection location. Data collection can happen at the client or server side of a communication channel, at the edge of the network, at the core of the network or any place in between. Collection point can dramatically affect available features, reliable labeling and generalization, which are discussed next. 

\subsubsection{Reliable Labeling}
Correct labels are crucial to the performance of traffic classification methods. However, labeling data is not always trivial. Some studies used free DPI modules, such as nDPI and libprotoident, to label captured data. In such cases, the accuracy of the labels, and thus any corresponding classification algorithms, is limited by that of the DPI methods. Furthermore, such methods do not usually work for encrypted traffic. 

A controlled environment at the client-side of the communication would be the easiest place to label the data. This solution is only practical when the capturing point is close enough to the data source to make sure that there is no other source to affect the labeling. Moreover, even in the fully controlled environment, it is not easy to distinguish and remove background traffic completely. It has been shown that $70\%$ of the smartphones' traffic is background traffic and only $30\%$ is directly related to the user interactions \cite{stober2013you}. Despite the limitations, capturing data of each class in a controlled environment has been the most commonly-adopted strategy in practice.

\subsubsection{Available Features} Useful information in packets are not always available. Packets captured at wireless links or cellular communications are encrypted at layer 2 and consequently useful upper-layer header fields are not in plain text. Furthermore, at some capturing points, such as a router in the center of an ISP, one may only capture one direction of a flow due to the asymmetric nature of routing in the Internet. Moreover, interarrival time may get distorted when traffic is aggregated, which is more severe at the core of ISPs. This phenomenon transforms the distribution of interarrival time and heavily depends on network conditions, traffic load, and time. Packet length may also change when the traffic passes through a tunnel, proxy, etc. Finally, all these changes also affect the statistical features obtained from the entire flow. Hence, a model trained on a dataset captured at one capturing point may not be as accurate when used at another capturing point.

\subsubsection{Representative Dataset}
A representative dataset should contain diverse and abundant samples from each class to avoid overfitting. It has been shown that the accuracy drops by as much as $26\%$ when OS/vendor is different in the training and test set \cite{aceto2017traffic}. Furthermore, a model may overfit to user-specific features rather than traffic-specific features if dataset contains interactions of only one or few users. It is also a big limitation on studies that captured the traffic generated by a script \cite{rezaei2018} which probably have more deterministic behavior. In general, a dataset captured further away from the client-side of the communication, for example at the core of an ISP where diverse traffic is observable, is less exposed to this issue. The best way to guarantee that the trained model on the dataset is representative is to test the model on a test set that comes from different device/user configuration than the training set.

%In \cite{balachandran2010volunteer}, they proposed a way to encourage volunteers to install an app and collect data from them. That anaylsed the sockets and applications that open them to label traffic.

\subsection{Dataset Pre-processing}
Data cleaning and pre-processing significantly affect the performance of ML algorithms. In a network environment, some relatively common events can change the packet-level  feature distribution. For instance, packet retransmissions, duplicate acks and out of order packets may change the traffic pattern of an application. Some studies reported improvement upon removing such packets \cite{dubin2017know} and some reported no difference \cite{aceto2017traffic}. That is because different datasets and features are used for classification. For example, methods that use statistical features of the entire flow are probably immune to a few unrelated packets. On the other hand, methods that use first few packets for classification might be affected more. Note that this pre-processing step is sometimes ignored due to its computational complexity.
%Moreover, it has been shown \cite{grimaudo2014select} that $90\%$ of flows are mice flows and have less than  6 packets which have been ignored by many studies. These flows only account for less than $1\%$ of the total traffic volume.

Another pre-processing step which is crucial to the performance of deep learning methods is data normalization. In this step, all input features are scaled to have a value in the range $[-1,+1]$ (or $[0,1]$). This allows the gradient-based methods to converge faster and equalizes the importance of all features when computing the distance between data points. 
%To normalize the features, one should know the minimum and maximum values which are not always trivial. For cases, where the raw packet bytes are taken as an input of the deep learning algorithm, it is reasonable to assume the maximum value is 255 for each byte and then rescale each byte. For statistical features, however, one may observe values greater than the maximum value observed in the training set. But, this is not usually problematic as far as the same rescaling is also applied to the test data. 
%For categorical data where they only take a limited number of values and the measure of distance does not represent similarity, such as QoS or protocol field in IP packet, one-hot encoding that uses a separate binary data type for each possible value often performs better. Nevertheless, these features are often ignored in traffic classification because they do not have much useful information, except for protocol field. 
%Since protocol field takes only 2 possible values (TCP and UDP) in most datasets and measure of distance gets important when there are more than 2 possible values, rescaling also works.

\subsection{Features}
%Deep learning methods can potentially deal with huge input dimensions by learning the most salient features automatically. However, in many practical applications, such as traffic classification at ISPs for QoS provisioning, it is desirable to classify with as few packets as possible, as few features as possible and as fast as possible. Hence, a method that rely on the observation of an entire flow for classification is not practical for such demands. The capability of online classification is determined by the application and available features. Moreover, online classifiers are often limited on the amount of pre-processing they can handle due to the limited time and computational capacity, and also on the amount of memory they access to.

State-of-the-art traffic classification methods use one or more categories of the features:

\textbf{Time Series:} Time series features include  packet length, inter-arrival time, and direction of consecutive packets. In many studies where these features were representative, the first few packets up to first 20 packets have been shown to be enough for reasonable accuracy even for encrypted traffic \cite{lopez2017network}. Times series features of a set of  sampled packets have also been recently shown to achieve good accuracy \cite{rezaei2018}.

\textbf{Header:} This includes all useful header fields in a packet, typically layer 3 and layer 4 information, when unencrypted. In pre-deep-learning era, fields including port number, protocol, and packet length, were carefully chosen by domain experts as representative features. In some recent approaches, especially deep-learning-based ones, entire packets are taken as input \cite{lotfollahi2017deep}. Note that server IP addresses might be used to limit the range of traffic classes for better accuracy in operational networks. For instance, one can use Google's IP addresses to limit traffic classes to Google's applications. However, IP addresses should be used judiciously due to the widespread use of CDNs and the dynamic allocation of IP addresses.

\textbf{Payload Data:}
Even for encrypted traffic, information above layer 4 header exists that can be exploited for classification. For instance, some studies have achieved high accuracy using TLS 1.2 handshake packets that contain  plain text data.

\textbf{Statistical Features:} There are numerous statistical features that can be obtained from the entire flow, such as average packet length, maximum packet length, and minimum inter-arrival time\footnote{ A comprehensive list of statistical features can be found at \url{https://centauri.stat.purdue.edu:98/netsecure/Papers/flowattributes_ademontigny.pdf}}. A large number of papers used these features and demonstrated high accuracy. However, to obtain statistical features, a classifier is required to  observe the entire or large portion of a flow and thus is only suitable for offline classification. Moreover, in some cases like application classification, statistical features can be affected by user-specific behaviors, OS-specific patterns, network-specific conditions, etc. Hence, dataset should be collected with more care.

Although time series and statistical features might be slightly different for unencrypted traffic and the encrypted version of the same traffic, they are available regardless of encryption. Hence, methods depending on these features for unencrypted traffic may also work with encrypted traffic as well. On the other hand, payload data and some header information, for instance layer 4 information of traffic encrypted by IPsec, might not exist in plain text for encrypted traffic. However, in these cases, there are still unencrypted fields available during handshake that can be used for classification. It is worth mentioning that in some cases privacy policies and laws prohibit accessing or storing packet content which limit the use of payload features.

\begin{figure*}
\centering
\subfloat[]{\includegraphics[width = 2in]{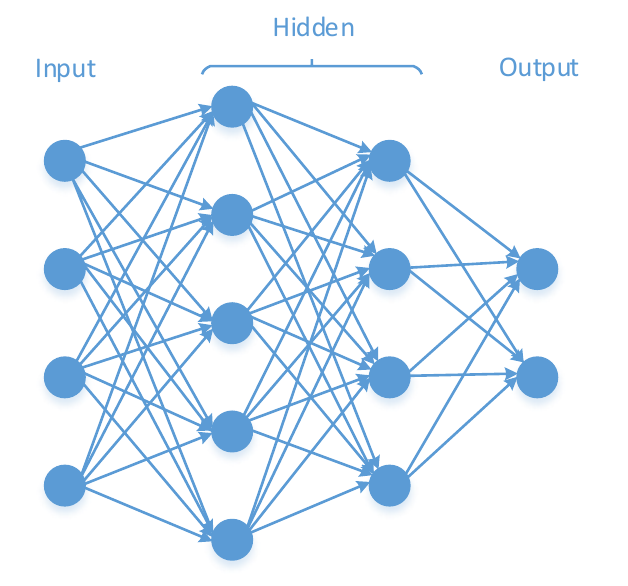}} 
\subfloat[]{\includegraphics[width = 2in]{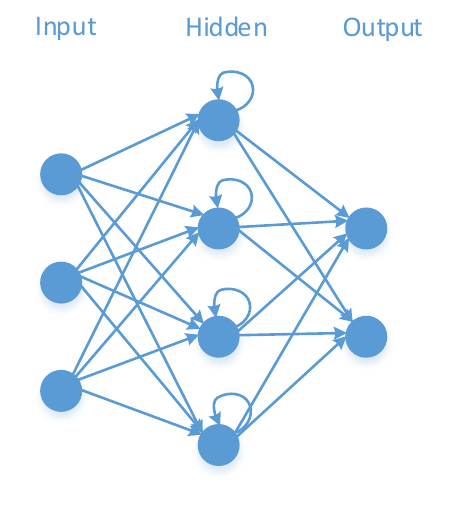}}\\
\subfloat[]{\includegraphics[width = 6in]{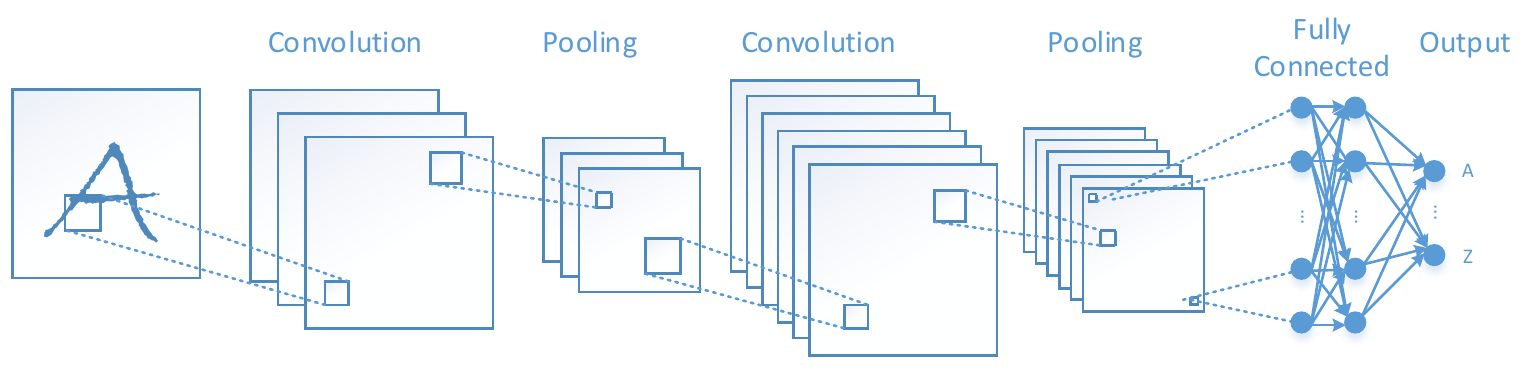}}\\
\subfloat[]{\includegraphics[width = 3in]{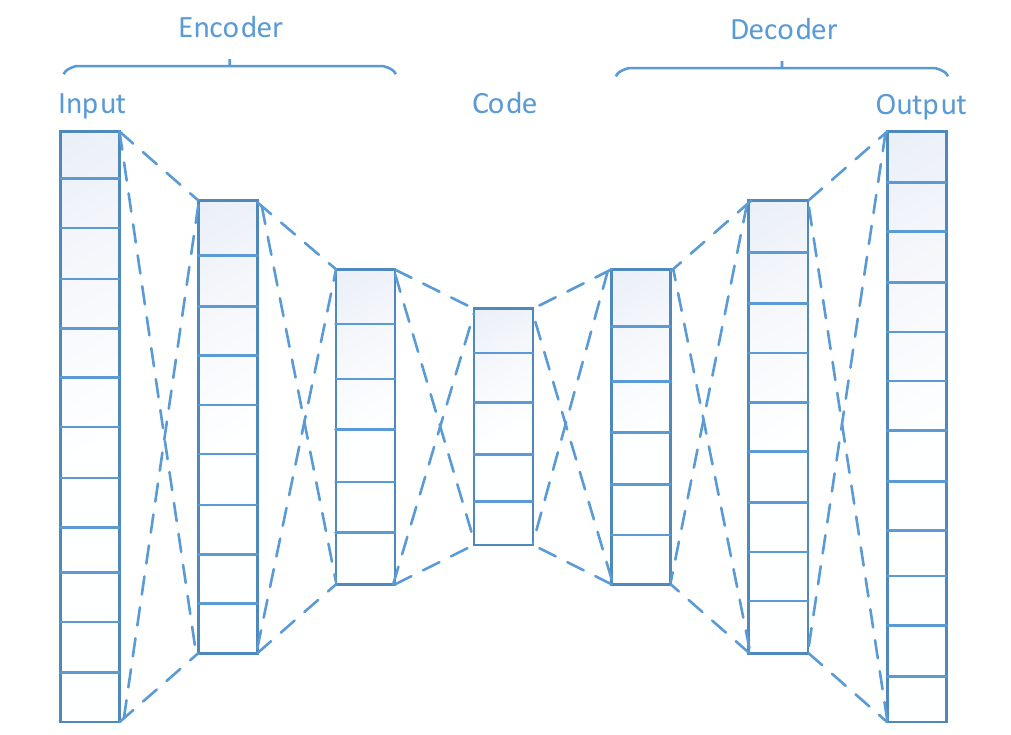}} 
\subfloat[]{\includegraphics[width = 3in]{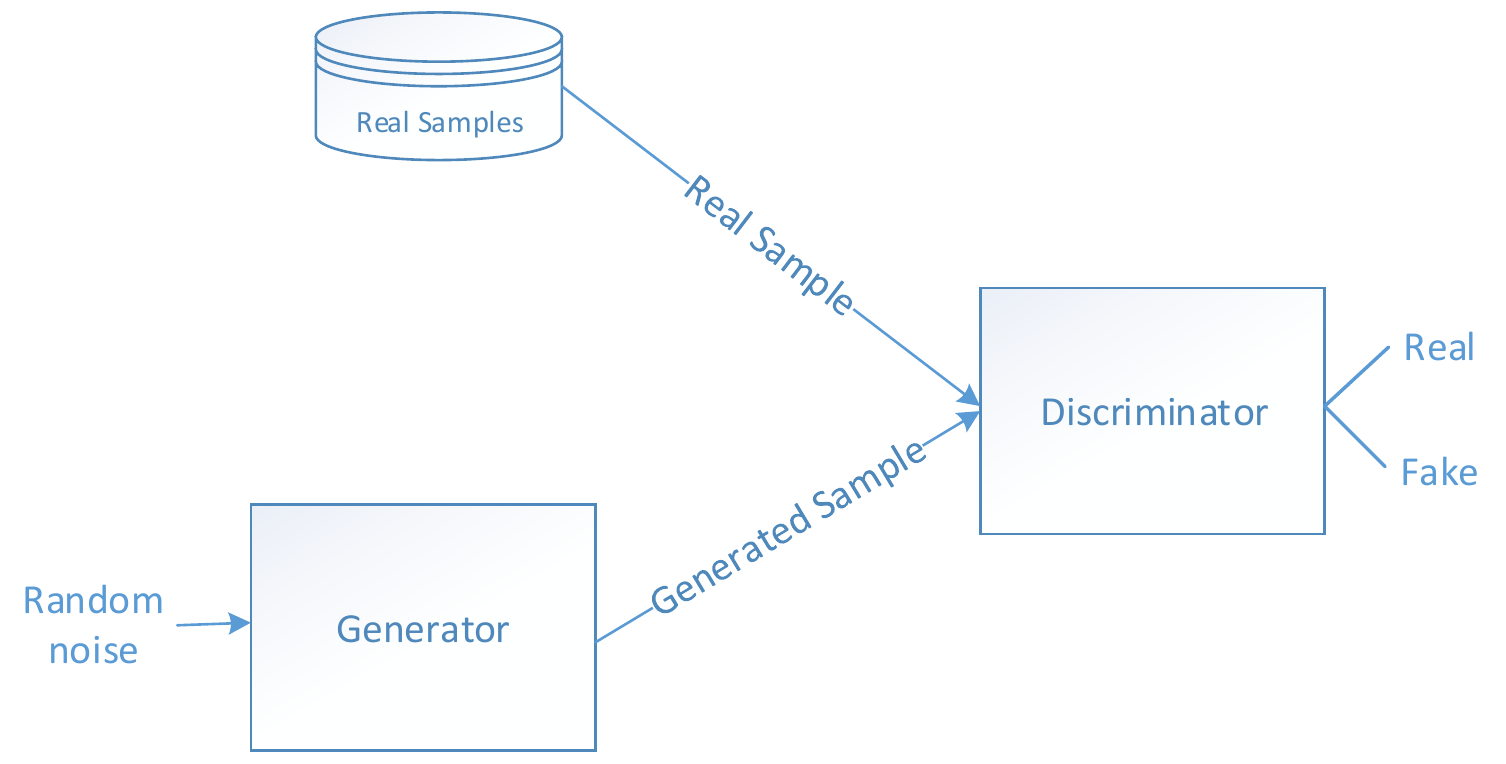}} 
\caption{Common deep learning models: a) MLP, b) CNN, c) RNN, d) AE, e) GAN.}
\label{fig-models}
\end{figure*}

\section{Deep Learning Techniques}

In this section, we briefly introduce some recent papers for each deep learning method, the summary of which is shown in Table I. 
%For unfamiliar readers, a brief introduction of deep learning methods can be found in \cite{fadlullah2017state}
Then, we complete our 7-step framework by explaining the model selection and evaluation in detail.

\subsection{Multi-Layer Perceptron}
Multi-layer perceptron (MLP) is the first neural network architecture and consists of an input layer, an output layer and  several hidden layers of neurons. Each layer has several neurons which are densely connected to adjacent layers, as shown in Fig. \ref{fig-models}(a). A neuron takes a weighted sum of its inputs and passes through a non-linear activation function to produce an output. Theoretically, a dense and deep enough MLP can estimate any arbitrary function. However, due to the huge number of parameters that a model need to learn, this model is usually very complex, inefficient and hard to train for an arbitrary complicated problem. Although the use of deep MLP alone has been declined, a few layers of fully-connected neurons, which can be considered as a MLP, is still used as small part of other models.

For network traffic classification, pure multi-layer perceptron (MLP) has been rarely used due to its complexity and low accuracy. In \cite{acetomobile}, many deep learning methods are compared with random forest (RF) algorithm to show the performance gap. They use 3 mobile datasets with different number of labels. Many deep learning methods outperform RF in two of the datasets. However, the experiment settings are not completely fair and equal because the input features used for RF, MLP and other deep learning methods are different. Hence, the results should not be considered as a comprehensive comparison of ML methods.
%For instance, RF classifier took statistical features as input while deep learning methods used either the first N bytes of the payload, or time series and header fields of the first 20 packets.

\subsection{Convolutional Neural Networks}

Similar to MLP, convolutional neural networks (CNNs) also consist of several layers with learnable parameters. MLP fails to work well with high dimensional input leading to a large number of learnable parameters in hidden layers. CNN architecture, shown in Fig. \ref{fig-models}(c), solves this problem by using convolution layers. In a convolution layer, a set of small kernels with a small number of learnable parameters are used. The same set of kernels are used on the entire input to produce the output for the next layer. By using the same set of kernels in a layer, the number of learnable parameters are dramatically reduced. The use of these kernels on the entire input helps the model to also capture shift invariant features more easily. Pooling layer is also used after one or few convolution layer for subsampling. Moreover, fully connected layers are usually used for the last hidden layers.

%CNN architecture showed significant improvement over traditional methods. In the last few years, many variations of CNN architecture have been introduced to allow deeper and more accurate models, such as GoogLeNet, ResNet and VGG.

The simplest convolutional neural network (CNN) model is proposed in \cite{wang2017end} which basically represent each flow or session with a 1-dimensional vector to feed the CNN model. Their CNN model has 2 convolutional, 2 pooling, and 2 fully connected layers. They normalize the bytes in each packet and use only the first 784 bytes. They evaluate their model on an encrypted application dataset of 12 classes and show a significant improvement over C4.5 approach that use time series and statistical features. In \cite{chen2017seq2img}, authors also use CNN with 2 convolutional, 2 pooling, and 3 fully connected layers for protocol and application classification tasks. They use reproducing kernel Hilbert space (RKHS) embedding and convert the early time series data into 2-D images. Their CNN model outperforms classical ML methods and MLP in protocol and application classification task. A semi-supervised approach based on a simple 1-dimensional CNN is used in \cite{rezaei2018} to classify five Google applications. They train a model to predict the statistical features of the entire flow from a few sampled packets with a large unlabeled dataset. Then, they transfer the weights to a new model and re-train it for application classification task with only a few labeled samples. They show the possibility of using sampled time series features instead of a first $n$ packets, which is more feasible for high bandwidth operational networks \cite{rezaei2018}.

\begin{table*}[t]
  \centering
  \label{table1}
  \begin{tabular}{*{15}{c}}
    \hline
    Paper & Category & DL method & Online & Features & Year \\
    \hline
    Wang-2018\cite{wang2018hast} & Intrusion detection & CNN+LSTM &  \ding{51} & Header+payload & 2018 \\
    \hline
    Rezaei\cite{rezaei2018} & APP/OS identification & CNN  &  \ding{53} & Sampled time series & 2018 \\
    \hline
    Aceto\cite{acetomobile} & APP classification & CNN/LSTM/SAE/MLP &  \ding{51} & Header+payload & 2018 \\
    \hline
    Vu \cite{vu2017deep} & Traffic identification & AC-GAN &  \ding{53} & Statistical  & 2017 \\
    \hline
    Wang-2017\cite{wang2017end} & Traffic identification & CNN  & \ding{51} &  Header+payload & 2017 \\
    \hline
    Seq2Img\cite{chen2017seq2img} & APP/protocol identification & RKHS+CNN  &  \ding{51} & Time series& 2017 \\
    \hline
    Lotfollahi\cite{lotfollahi2017deep} & APP/traffic identification & CNN/SAE  &  \ding{53} & Header+payload & 2017 \\
    \hline
    Lopez-Martin\cite{lopez2017network} & Mixed-type classification & CNN+LSTM  &  \ding{51} & Header+time Series & 2017 \\
    \hline
    Hochst\cite{hochst2017unsupervised} & Traffic identification & Autoencoder  &  \ding{53} & Statistical+header & 2017 \\
    \hline
  \end{tabular}
  \caption{Overview of deep learning methods used for traffic classification.}
\end{table*}

\subsection{Recurrent Neural Networks}

Recurrent neural networks (RNNs) are neural networks containing loops to store temporal information, as shown in Fig. \ref{fig-models}(b). RNNs were designed specifically for sequential data where the output may depend not only on the last input but on the previous inputs as well. RNNs have been successfully applied to speech recognition, time series prediction, translation, and language modeling. Gradient vanishing and exploding, which makes learning long-term dependencies difficult (e.g. dependencies between inputs that are far apart), was a common obstacle in traditional RNNs. The long short-term memory (LSTM) was introduced to mitigate these problems by adding a set of gates that control when information is stored or removed. 

For network classification tasks, mixed models are reported to outperform pure LSTM or CNN models \cite{lopez2017network}. To capture both spatial and temporal features of a flow, both CNN and RNN are used in \cite{wang2018hast,lopez2017network} for different applications. Aside from minor differences, both studies take the content of the first 6 to 30 packets to the CNN model followed by a RNN or LSTM model. Although the exact input features, the neural network architectures, and the datasets are different, they both report high accuracy.

Despite their success in sequential data, LSTMs are not suitable for complex tasks which requires explicit and external memory. New architectures, such as memory networks and neural Turing machines (NTMs), have been recently introduced to embed explicit memory into the architecture, referred to as memory augmented neural networks (MANN). MANNs have been successfully applied in language modeling, question answering and one-shot learning. The performance of MANNs on the network classification task has not been studied yet.

%\subsection{Memory-augmented Neural Networks}

\subsection{Auto-encoders (AE)}

Auto-encoders (AEs) are neural networks with significantly smaller hidden layers compared to the input and output that aims to reconstruct the input at the output, as shown in Fig. \ref{fig-models}(d). The internal encoded representation can be used for data compression or dimensionality reduction. MLPs, CNNs and RNNs can all be used as a part of the AE architecture. AE is extensively used to initialize the weights of deep architectures. There are some variations of AE, such as denoizing auto-encoders (DAEs) that are trained to output intact input samples by taking a corrupted samples forcing the model to learn more robust features, and variational auto-encoders (VAEs) that aim to generate virtual examples from a target distribution. More complex architectures, called stacked auto-encoders (SAEs), stack up several AEs where the output of each one is the input of the next AE and the whole model is trained in a greedy layer-wise fashion. It is also possible to train a model, called hybrid learning framework, which combines AE with MLP, or other models, with labeled data from the beginning. Hence, the model learns both the input and output distribution at the same time. Such hybrid models are trained with multi-objective loss functions including standard output losses as well as the input reconstruction terms.

AEs are usually used in an unsupervised fashion to obtain smaller representation of input data which can be later used as a part of a classifier. For instance, in \cite{hochst2017unsupervised}, an auto-encoder is used to reconstruct the input. Then, a softmax layer is applied to the encoded internal representation of the auto-encoder and they achieve a moderate accuracy. They use their own private dataset with 7 traffic types. Moreover, they use nine statistical features of 12 intervals and both flow directions as an input. In \cite{lotfollahi2017deep}, authors take header and payload data to train a 1-D CNN and a SAE model on ISCX VPN non-VPN dataset. Both models show high accuracy, but the CNN model marginally outperforms the SAE model.

\subsection{Generative Adversarial Networks}

Generative adversarial network (GAN) is an unsupervised technique that trains a generative and a discriminative model simultaneously. As shown in Fig. \ref{fig-models}(e), the generator aims to generate (fake) examples of the target distribution and the discriminator model aims to distinguish between real and generated data. Both models are usually neural networks. The generator is first trained to maximize the error probability by discriminator. Then, the generator is fixed and discriminator is trained to minimize the error probability while real and generated data is fed in. The procedure is continued until it converges. Despite the difficulty of training and converging GANs, it has been used in many applications, such as creating realistic images, reconstructing 3D models from images, improving image quality, creating synthesized data for applications with scarce data.

Generative models can be used to handle dataset imbalance problem in network traffic classification. The imbalance problem refers to scenarios where the number of samples for each class varies considerably. In such cases, ML algorithms usually have difficulties predicting the minority classes correctly. The most frequent and easiest approach to deal with imbalance dataset is oversampling minority classes, duplicating samples of minor classes, or undersampling majority classes, removing some samples from major classes.
In \cite{vu2017deep}, auxiliary classifier GANs (AC-GAN) is used to generate synthesized samples for supervised network classification task. The main difference between AC-GAN and GAN is that AC-GAN takes both a random noise and a class label as input so as to generate the sample of the input class label. They use a public dataset with two classes, SSH and non-SSH, and 22 statistical features for the classifier input. They use deep models only to generate synthesized data. For classification part, they use classical ML algorithms, including SVM, RF, and decision tree.

\subsection{Model Selection}

Several factors affect the choice of deep learning models for network traffic classification. The most important one is the choice of features. Features directly affect input structure and dimension which influence computational complexity and number of packets for classification (memory complexity). Next, one should choose a suitable model based on the chosen feature. Here, we do not cover header features alone and we only cover it in conjunction with other features because header features alone are not always effective enough for classification. In headers, only port number, window size, and in some rare cases type of service (ToS) or fragmentation-related fields provides information useful for classification. 

 The choice of input feature and machine learning method are highly correlated. Furthermore, the size of the dataset also affects the choice of the model. For instance, deep learning methods are not suitable when the dataset is small. Assuming that the dataset is large, three commonly used input features and the corresponding suitable models are described below:

\textbf{Time Series+Header:} Since time series features are barely affected by encryption, it has been widely applied to various applications and datasets. The first few packets, from 10 to 30 packets, are reported to be enough for classification in many datasets \cite{chen2017seq2img,lopez2017network}. Sampled packets from the entire flow are also shown to achieve promising accuracy \cite{rezaei2018}. Classical ML algorithms and MLP models work well when the number of packets, representing the input dimension, is small. For a larger number of packets, CNN and LSTM are reported to be more accurate 
\cite{lopez2017network}. However, even for small number of packets, CNN model is used \cite{lopez2017network,rezaei2018}. But, in general, computational complexity and training time of deep models are higher than classical machine learning algorithms. 

\textbf{Payload+Header:} In current encrypted traffics, the first few packets that contain handshake information are typically unencrypted and they have been successfully used for classification. Due to the high dimensionality of the input (large number of bytes in payload), classical ML methods and MLP do not work well. In such cases, CNN or combination of CNN and LSTM are reported to have high accuracy \cite{wang2017end},\cite{acetomobile},\cite{lotfollahi2017deep},\cite{wang2018hast}. It is possible to also use time series features alongside payload information to slightly improve accuracy, but this barely changes the input dimension or the choice of model.

\textbf{Statistical Features:} The number of statistical features, and consequently the input dimension, is limited. Hence, most papers used classical ML methods or in rare cases MLP for these features. Although most studies obtained statistical features by observing the entire flow, it has been shown that obtaining statistical features from the first 10 to 180 packets, depending on the datasets and the choice of statistical features, might be sufficient for classification.  Even though statistical features allow us to build a simpler classifier based on classical machine learning algorithms, it may not be suitable for online and fast classification because it needs to capture enough packets to obtain dependable statistical features from a flow.

Table II summarizes features, the corresponding models, and their properties. Note that there is no guarantee that all these approaches work for a particular dataset. That is the reason why one might need to go to data collection step if data is not enough, or to feature or model selection step if chosen features or model are not representative\footnote{ We haven't shown direct arrow from training and validation step to feature and model selection steps since one can simply skip the unnecessary steps starting from data collection.}. Moreover, these approaches have only studied on certain traffics. Comprehensive study of ever-increasing and upcoming protocols, such as QUIC and TLS 1.3, has not been done yet.

\begin{table*}[t]
  \centering
  \label{table2}
  \begin{tabular}{*{15}{c}}
    \hline
    Feature & Time series+header & Payload+header & Statistical \\
    \hline
    Model & Classical ML/MLP/CNN/LSTM & CNN/CNN+LSTM & Classical ML/MLP \\
    \hline
    Computational complexity & Low/Medium &  High & Low \\
    \hline
    Number of packets needed & Medium &  Low & High \\
    \hline
  \end{tabular}
  \caption{Guide for model and feature selection}
\end{table*}

\subsection{Training and Validation}
Training and validation step is similar to any other deep learning applications where a model's hyper-parameters are tuned to obtain the best accuracy. Typically, dataset is divided into three separate sets: train, validation and test set. The model is trained on the train set and the accuracy of validation set is observed to tune model's hyper-parameters. Finally, the unbiased accuracy is obtained by using the test set. The detailed best practices of the last two steps are outside the scope of this article. One can read a training and validation guideline on any other application and apply the same best practices here. 

\subsection{Periodic Evaluation/Update}

The last step, periodic evaluation/update, has not been comprehensively studied yet. In most network-related applications, traffic characteristic of classes is always changing. Moreover, new traffic classes, called zero-day applications, are constantly emerging. Only a limited number of papers studied such challenges and they are still open problems worth more comprehensive analysis. They are briefly discussed in open problems and opportunities section.

\section{Open Problems and Opportunities}

In traffic classification, unencrypted traffic classification has been extensively studied and many commercial and free tools have been developed for.  Classification of encrypted traffic is a harder task due to the lack of representative features, but a few studies have shown successful classification of TLS 1.2 and VPN traffic in UDP mode. It is still not clear whether these methods can handle significantly larger number of  classes common in operational network. There are also many unsolved problems in traffic classification that we will introduce in this section.

\subsection{Stronger Encryption Protocols}
Traffic classification for stronger encryption protocols, in particular QUIC and TLS 1.3, has not been well investigated. Most browsers, including Chrome and Firefox, have already implemented TLS 1.3 draft version. However, most applications and websites have not adopted TLS 1.3 yet.

Previous studies on TLS 1.2 mainly used plain text fields during the handshake. But, by the introduction of 0-RTT connectivity in TLS 1.3 and QUIC, only a few fields in the first packet remain unencrypted which is not clear if they suffice for classification.  In \cite{rezaei2018}, QUIC protocol is classified using sampled time series features. But, the classification is limited to five Google applications and the accuracy was high when only training and test set is captured based on script-generated traffic.

\subsection{Multi-label Classification}
A single flow may contain more than one class label, referred to as multiplexed stream. For instance, a traffic that passes through a tunnel may contain several applications that share the same 5-tuple. QUIC protocol also may contain several classes of traffic. There is no method in traffic classification or related literature to deal with these cases. The most first and most difficult challenge is how to collect and label such traffic appropriately.

\subsection{Middle Flow Classification}
Around $90\%$ of the flows are short-lived ones. In certain applications, such as traffic engineering, one may want to focus on long flows. However, if the classification method relies on the first few packets, an ISP should store the first few packets of all flows which are huge burden. On the other hand, if the classification method works with packets in the middle of the flows, ISPs can wait and detect elephant flows, and then classify the elephant flows by capturing a few packets from the middle of the flow. This will dramatically reduce the memory and computational overhead. A few studies have shown that the accuracy is higher when the first few packets are involved in classification, but no comprehensive study has conducted to use a set of packets from an arbitrary point in the middle of the flow. Note that some studies divide the entire flow into several bursts and then classify each burst to detect different user actions \cite{taylor2018robust}. This means that the beginning of the burst should also be detected and the capturing process must be started at this specific point. Moreover, it is not clear whether this method also works for other classification problem rather than user action. In \cite{rezaei2018}, authors used fixed number of sampled packet from different part of a flow for classification. They achieved a moderate accuracy when sampling from anywhere in the flow. However, high accuracy from middle flow is still an open problem.

\subsection{Zero-day Applications}
Zero-day applications refer to the traffic classes that are new and their samples do not exist in the training set. It has been shown that in some cases zero-day applications can make up to $60\%$ of flows and $30\%$ of bytes in a network traffic \cite{zhang2015robust}. Despite the importance, it is in a nascent stage and only a few recent studies \cite{zhang2015robust} proposed solutions which usually rely on detecting unlabeled clusters and then labeling them. In the ML community, active learning, where a model selects which data points should be labeled, has been studied for many years. In a recent study on classifying images of characters \cite{woodward2017active}, a combination of reinforcement learning and LSTM is used to perform one of the two possible actions: predicting the class or asking for a new label. There are many useful ideas in the ML community that can be adopted to solve the zero-day application problem.

%\subsection{App in App Classification}

\subsection{Transfer Learning and Domain Adaptation}

It is not always possible to collect a large enough representative dataset. It is often easier to obtain large datasets captured for other tasks, which may help the model to extract common features. Moreover, training a deep model usually takes from a few hours to a few days or weeks, depending on the model size and dataset.  Since retraining a model often converges faster, it is preferable to retrain a model that has already been trained for similar task. Transfer learning and domain adaptation are the two widely used techniques in ML to achieve such a goal.

Transfer learning allows a model trained on a source task to be used on a different target task. The assumption is that the input distribution of the source and destination tasks are similar. This process only works when the features learned by the model are not specific to the source task. Since the model is already trained to capture useful features, the retraining process on the target task needs significantly less labeled data and training time. In the case of network traffic classification, a publicly available dataset can be used to pre-train a model which will be later tuned for another traffic classification task with fewer labeled samples. 

A recent paper \cite{rezaei2018} used this approach by transferring the weights of a pre-trained CNN model to a new model that later was trained to classify Google applications. The model was first trained to predict statistical features of the entire flow from sampled packets, which does not need human effort for labeling. Then, the model was transferred and re-trained with a small labeled dataset containing only 20 labeled samples for each class. They also showed that the model pre-trained on a public unrelated dataset can still be used for transfer learning. But, the accuracy of that scenario was limited to below $85\%$.

%In classic ML tasks, such as image classification, transfer learning has shown to be highly effective. It is usually done in two ways: (a) One common strategy is to train a model on a similar but previously labeled dataset and then tune the model on the target task, hopefully with less labeled data and faster convergence. This approach has also been used in k-shot learning methods where only k labeled samples are available for the target task. (b) The other strategy is to train a model on an unlabeled dataset, by constructing the same input with a generative model, such as auto-encoder, and then retrain the model on the target task.

Unlike transfer learning where the source and target tasks (i.e. their class labels) are different, domain adaptation deals with the cases where the task is the same, but the input distribution of the source and target is different. Although it is different from transfer learning, similar techniques have been used to solve both problems. An example in the context of network traffic classification would be to train a traffic classifier model with a dataset captured at client side of the communication and then adopt the model to classify traffic at the core of the network where the data distribution is different. Another example is the case in which one can re-train a model periodically based on domain adaptation techniques to capture new patterns for classes whose features are constantly changing, which are not uncommon in current Internet. Despite their usefulness, these strategies have not been extensively adopted for network classification task yet.

\subsection{Multi-task Learning}

This approach refers to any model in which more than one loss function is being optimized. One typical approach is to share the hidden layers among all tasks, while each task has its own output layer. It has been shown that it reduces the risk of overfitting and helps the model find relevant features faster. This works when the input data is generated from a similar probability distribution or can be generated using a set of transformations from one another. As a result, it may be possible to use additional available datasets and define a single task for each if they are similar to your target task dataset. This can easily augment the dataset and improve the generalization. Many variations of multi-task learning have been used successfully for natural language processing and computer vision.

It has been shown that even for single-task problems, adding some auxiliary task will improve generalization and performance. However, it has not been studied for network traffic classification task. There are potentially many ways to define auxiliary task without the need for additional labeling. For instance, assume a typical model which takes time series information of the first 20 packets as an input. One can define many auxiliary tasks that do not need human labeling, such as detecting TCP/UDP class, predicting the average packet length of the entire flow, detecting mice/elephant flow, etc. The efficacy of multi-task learning has not been studied for network traffic classification yet.

% Can use something like this to put references on a page
% by themselves when using endfloat and the captionsoff option.
\ifCLASSOPTIONcaptionsoff
  \newpage
\fi

% trigger a \newpage just before the given reference
% number - used to balance the columns on the last page
% adjust value as needed - may need to be readjusted if
% the document is modified later
%\IEEEtriggeratref{8}
% The "triggered" command can be changed if desired:
%\IEEEtriggercmd{\enlargethispage{-5in}}

% references section

% can use a bibliography generated by BibTeX as a .bbl file
% BibTeX documentation can be easily obtained at:
% http://mirror.ctan.org/biblio/bibtex/contrib/doc/
% The IEEEtran BibTeX style support page is at:
% http://www.michaelshell.org/tex/ieeetran/bibtex/
%\bibliographystyle{IEEEtran}
% argument is your BibTeX string definitions and bibliography database(s)
%\bibliography{IEEEabrv,../bib/paper}
%
% <OR> manually copy in the resultant .bbl file
% set second argument of \begin to the number of references
% (used to reserve space for the reference number labels box)

\nocite{*}
\bibliographystyle{IEEEtran}
%\bibliography{References}

% biography section
% 
% If you have an EPS/PDF photo (graphicx package needed) extra braces are
% needed around the contents of the optional argument to biography to prevent
% the LaTeX parser from getting confused when it sees the complicated
% \includegraphics command within an optional argument. (You could create
% your own custom macro containing the \includegraphics command to make things
% simpler here.)
%\begin{IEEEbiography}[{\includegraphics[width=1in,height=1.25in,clip,keepaspectratio]{mshell}}]{Michael Shell}
% or if you just want to reserve a space for a photo:

\begin{IEEEbiography}[{\includegraphics[width=1in,height=1.25in,clip,keepaspectratio]{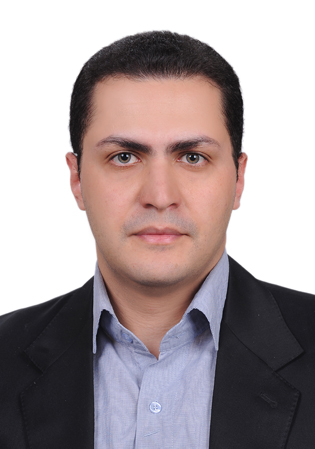}}]{Shahbaz Rezaei}
received his B.S. degree in Computer Engineering from University of Science and Culture, Tehran, Iran, in 2011 and M.S. degree from the Sharif University of Technology, Tehran, Iran, in 2013. His research interests include computer networks, performance modeling, machine learning and deep reinforcement learning. He is currently a Ph.D. student at UC Davis.
\end{IEEEbiography}

\begin{IEEEbiography}[{\includegraphics[width=1in,height=1.25in,clip,keepaspectratio]{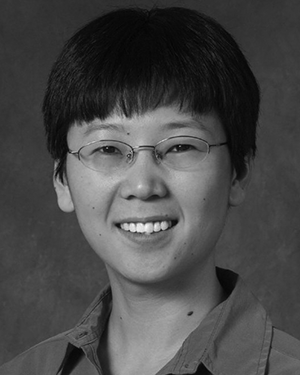}}]{Xin Liu}
received the Ph.D. degree in electrical
engineering from Purdue University in 2002.
She is currently a professor in the Computer Science Department, University of California, Davis, CA, USA. Before joining UC Davis, she was a postdoctoral research associate in the Coordinated Science Laboratory at UIUC. From March 2012 to
July 2014, she was on leave from UC Davis and with Microsoft Research Asia. Her research interests are in the area of wireless communication networks,
with a current focus on data-driven approach in networking. Dr. Liu received the Best Paper of year award of the Computer Networks Journal in 2003 for her work on opportunistic scheduling. She received the NSF CAREER award in 2005 for her research on Smart-Radio-Technology-Enabled Opportunistic Spectrum Utilization, and the Outstanding Engineering Junior Faculty Award from the College of Engineering, University of California, Davis, in 2005. She became a Chancellor's Fellow in 2011.
\end{IEEEbiography}

% You can push biographies down or up by placing
% a \vfill before or after them. The appropriate
% use of \vfill depends on what kind of text is
% on the last page and whether or not the columns
% are being equalized.

%\vfill

% Can be used to pull up biographies so that the bottom of the last one
% is flush with the other column.
%\enlargethispage{-5in}

% that's all folks
\end{document}